\documentclass[preprint,showpacs,preprintnumbers,amsmath,amssymb]{revtex4}
\usepackage{graphicx}
\usepackage{dcolumn}
\usepackage{bm}

\begin{document}


\title{Direct evidence of spin-split  one-dimensional metallic surface state on Si(557)-Au
}

\author{Taichi Okuda$^1$}
\email{okudat@hiroshima-u.ac.jp}

\author{Koji Miyamaoto$^1$}%

\author{Yasuo Takeichi$^2$}%

\author{Hirokazu Miyahara$^3$}%

\author{Manami Ogawa$^2$}%

\author{Ayumi Harasawa$^2$}%

\author{Akio Kimura$^3$}%

\author{Iwao Matsuda$^2$}%

\author{Akito Kakizaki$^2$}%

\author{Tatsuya Shishidou$^{4,5}$}%

\author{Tamio Oguchi$^{4,5}$}%

\affiliation{%
$^1$Hiroshima Synchrotron Radiation Center (HSRC), Hiroshima University, 2-313 Kagamiyama, Higashi-Hiroshima 739-0046, Japan
}%
\affiliation{%
$^2$The Institute for Solid State Physics (ISSP), The University of Tokyo, 5-1-5 Kashiwanoha, Kashiwa 277-8581, Japan
}%
\affiliation{
$^3$Graduate School of Science, Hiroshima University, 1-3-1 Kagamiyama, Higashi-Hiroshima 739-8526, 
Japan 
}%
\affiliation{
$^4$Department of Quantum Matter, ADSM, Hiroshima University, 1-3-1 Kagamiyama, Higashi-Hiroshima 739-8530, 
Japan 
}%
\affiliation{
$^5$ADSM\&IAMR, Hiroshima University, 1-3-1 Kagamiyama, Higashi-Hiroshima 739-8530, 
Japan 
}%

\date{\today}

\begin{abstract}
We report unprecedented evidence of a spin split  one-dimensional metallic surface state for the system of Si(557)-Au obtained by means of high-resolution spin- and angle-resolved photoelectron spectroscopy combined with first principles calculations. The surface state shows double parabolic energy dispersions along the Au chain structure together with a reversal of the spin polarization with respect to the time-reversal symmetry point as is characteristic from the Rashba effect. Moreover, we have observed a considerably large out-of-plane spin polarization which we attribute to the highly anisotropic wave function of the gold chains.

\end{abstract}

\pacs{Valid PACS appear here}
\maketitle

Low dimensional systems are well known as sources for unprecedented and intriguing new physical phenomena. These effects arise when electron confinement and symmetry breaking effects become dominant. If strong spin-orbit interaction is added to this list, then a momentum dependent spin-splitting of the electronic states is likely to happen. This phenomenon, initially derived from the relativistic Dirac equation, is commonly known as the Rashba effect and its Hamiltonian is written as \\
\begin{equation}
H_{soi}=-\frac{\hbar^2}{4m^2c^2}(\nabla V\times\overrightarrow{p})\cdot\overrightarrow{\sigma},
\end{equation}
where $V$ is the electronic potential, $\overrightarrow{p}$ the momentum operator and $\overrightarrow{\sigma}$ the vector of Pauli matrices.
Unlike the Zeeman splitting, the Rashba spin splitting does not result in a finite global spin-orientation in two-dimensional systems.
On the other hand, a global spin projection is expected for one-dimensional systems ---where electron propagation is restricted to one spatial direction---, as theoretically predicted for thin cross-sectional nanowires \cite{Governale04}.

An experimental demonstration of a one dimensional Rashba effect is yet unavailable. Model systems for this would be Au chains grown on vicinal silicon surfaces such as Si(335), Si(553), Si(557) and Si(775) \cite{Crain04}. Particularly, Si(557)-Au appears to be an excellent candidate since its structure has been well established both by X-ray diffraction\cite{Robinson02} and by density functional theory (DFT) calculations\cite{DSP02}. The resulting structure is illustrated in Fig. 1 and shows its main building blocks: Au atomic chains ($\gamma$), which grow at the center of the terraces, Si adatoms ($\alpha$), having a double ($\times$2) periodicity, and Si honeycomb chains ($\eta$). The valence band electronic structure of this system shows an evident one-dimensional metallic character with parabolic  energy dispersion along the chain direction and almost straight Fermi surface lines perpendicular to them, as observed in angle-resolved photoelectron spectroscopy (ARPES)\cite{Crain04}. 

Especial attention ---and a strong controversy--- has been generated by the experimental finding that the parabolic bands were doubled. At first, these double bands in Si(557)-Au were interpreted as a spin-charge separation in a Tomonaga Luttinger liquid\cite{Segovia99}. Subsequent high-resolution ARPES measurements\cite{Losio01} ruled out this hypothesis after revealing that the two bands do not merge at the Fermi energy. 
A later first-principles calculation proposed this splitting to be caused by a large spin-orbit interaction, in analogy to the case of the two-dimensional spin split surface state of gold\cite{DSP04}.
Other groups interpreted the origin of these electronic states based on purely structural mechanisms after the experimental finding by temperature controlled ARPES of differences in the electronic character of each band (one showing a metal-insulator transition at 78 K while the other being insulating (semiconducting) already at room temperature) \cite{Ahn03}. Furthermore, dimerization of the step-edge Si atoms was reported by low temperature scanning tunneling microscopy at the transition temperature \cite{Yeom05} which could naturally explain each surface state as originating from different surface structures.

However, some recent works have brought back the Rashba type spin splitting scenario: First, a recent high-resolution ARPES study on the related system of Si(553)-Au, showed the characteristic anti-crossing of the surface states at $\times$2 zone boundary\cite{Barke06}. Second, the observation of a particular surface plasmon on Si(557)-Au points towards a surface state which is  spin split\cite{Nagao06}.

It is evident that this controversy of the doubled surface state can only be settled by providing direct evidence of the spin structure of the system. Therefore, in this letter we present high-resolution spin- and angle resolved photoelectron spectroscopy (SARPES) measurements of the surface states of Si(557)-Au. Given that standard Mott type spin detectors are so inefficient and hamper the high-energy and angular resolved ARPES measurements, for this system we have used  our newly developed SARPES system equipped with home-made high-efficiency spin detectors \cite{Okuda08} which are better suited to the study of surface states with very steep dispersion and small shift in $k$-space. In this way, we observe that the surface states show unambiguous spin splitting originating from a strong spin-orbit interaction (Rashba effect). The observed spin-polarization has not only an in-plane but also an out-of-plane component which, from our first-principles calculations, we attribute to the spatial distribution of the wave function of Au.

Si(557)-Au surface was prepared by deposition of 0.2 ML Au onto the atomically clean and well ordered Si(557)  surface\cite{Viernow98} at a substrate temperature of 650 $^{\circ}$C with a post-annealing at 850$^{\circ}$C \cite{Crain04}.
The Au coverage was calibrated from a time extrapolation
of the reconstructions of Si(111)$\sqrt{3}\times\sqrt{3}$-Au ($\theta$=0.67 ML) and Si(111)5$\times$2-Au ($\theta$=0.4 ML). 
The quality and  cleanliness of these samples were checked by reflected high energy electron diffraction, low energy electron diffraction, x-ray photoemission and Auger electron spectroscopy.
During measurements the surface was refreshed by short flashes at 850$^{\circ}$C \cite{Losio01} and  checked afterwards by gauging the quality of the surface states.

ARPES and high-resolution SARPES measurements were performed at beamlines BL-18A and BL-19A of the Photon Factory(PF), KEK, Japan. 
The spin-{\it integrated} ARPES measurement was done using a  SCIENTA SES-100 analyzer, whereas for SARPES measurement we used a home-made high-efficient very-low-energy-electron-diffraction (VLEED) type spin detector combined with a large hemispherical photoelectron energy analyzer (SPECS GmbH. PHOIBOS-150). 
The high efficiency of our new spin detector makes the high-energy and angular resolutions compatible with spin detection\cite{Okuda08}.
In this work we achieved  energy and angular resolutions of $\sim$60 meV and  $\pm$0.2$^{\circ}$ for the ARPES and $\sim$100 meV and $\pm$1$^{\circ}$ for the SARPES measurements, respectively.
The effective Sherman function of the spin detector was 0.33 as determined  from  Ni(110) \cite{Okuda08}.
As illustrated in Fig.1, the angle between the incident synchrotron beam and the analyzer entrance lenses was fixed at 45$^{\circ}$.

Both the ARPES and SARPES spectra have been acquired along Au chain structure, x$_{\rm s}$-direction, which corresponds to the dashed line shown in the surface Brillouin zone (SBZ)\cite{SBZ} (Fig.1). This was done by rotating the sample against the y = y$_{\rm s}$ direction using a photon energy h$\nu$=34 eV.
Since our spin-detector can measure transverse spin ($P_{\rm x}$: polarization along the x-direction of the spin-detector) and longitudinal ($P_{\rm y}$: y-direction) spin components, $P_{\rm x}$ contains a variable contribution of $P_{\rm z_{\rm s}}$ and $P_{\rm x_{\rm s}}$ components when the sample is rotated around the y = y$_{\rm s}$ axis.

\begin{figure}
\includegraphics[scale=0.3]{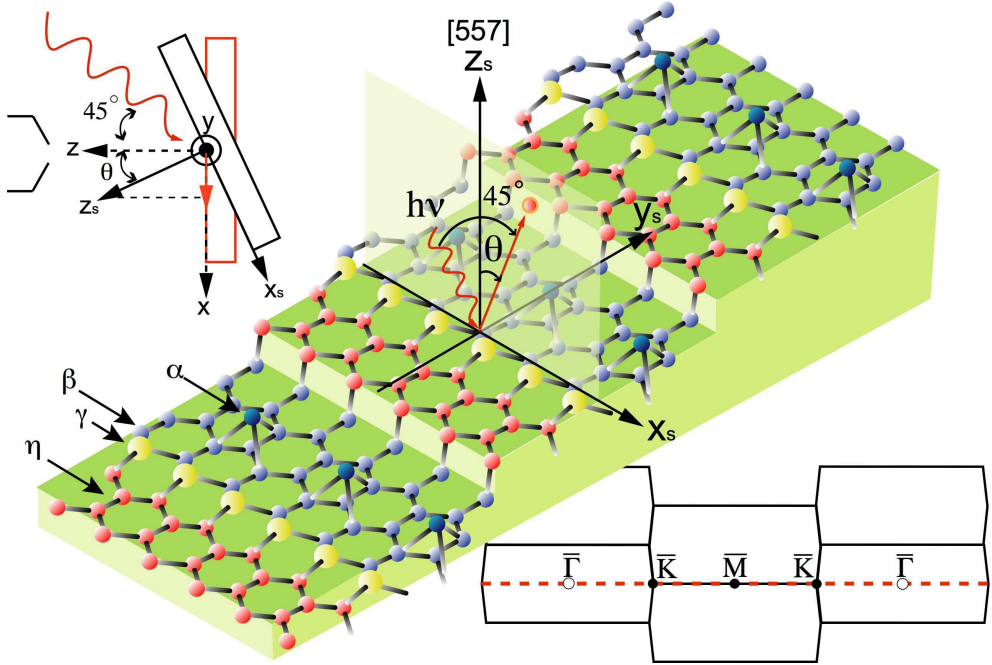}
\caption{\label{fig1:epsart} (Color online) Schematic illustration of Si(557)-Au model and the surface Brillouin zone (SBZ) as well as the experimental geometry. The atoms labeled $\alpha$ (dark blue), $\gamma$ (yellow) and $\eta$ (red) correspond to Si adatoms, Au atoms, and honey-comb structured Si atoms, respectively. (S)ARPES spectra have been acquired along gold chain direction (x$_{\rm s}$) which corresponds to the red dashed line in the SBZ. }
\end{figure}
%


Figure 2(a) shows a color scale map of the spin-integrated photoemission intensity vs $E$ and $k_{\parallel}$. These ARPES data were acquired at $T$=150 K around the $\bar{\rm M}$ point of the second SBZ. As has been previously reported, well resolved double parabolas which cross the Fermi energy ($E_F$) are observed  suggesting a metallic character of both states.
In our measurement we did not observe a clear indication of an energy gap in the $S_2$ state as reported in Ref. \cite{Ahn03}.
The surface state pair is only visible within the photon energy range of 26 eV to 40 eV and shows to be  most prominent at h$\nu$=34 eV. The photoemission intensity is also  angle dependent, such that stronger surface state intensities are recorded at higher emission angles, as previously reported \cite{Losio01,Altmann01}. 

Figure 2(b) shows a color scale map of the second derivative of the total intensity of up- and down-spin states ($I^{\uparrow}$+$I^{\downarrow}$)  vs $E$ and $k_{\parallel}$ obtained from the SARPES measurement.
Because of the lower energy and angular resolutions in SARPES compared to ARPES, the double parabola is smeared into a single one.
Selected SARPES spectra and its polarization for longitudinal ($P_{\rm y}$) and transverse ($P_{\rm x}$) spin detector directions are presented in Fig.2(c) and (d). The up-spin states (red thick lines) correspond to y and x directions in Fig. 1 for $P_{\rm y}$ and $P_{\rm x}$, respectively, having opposite sense for the down-spin (blue thin lines) states. Spin polarization is present in both longitudinal and transverse directions: Energetically shifted up- and down-spin states are clearly observed at $k_{\parallel}$=$\sim$1.05\AA$^{-1}$, $\sim$1.09\AA$^{-1}$, and $\sim$1.14\AA$^{-1}$.
The spin split states disperse towards the Fermi level showing the crossings at $k_{\parallel}$=$\sim$1.18\AA$^{-1}$ for the down-spin state and at $\sim$1.27\AA$^{-1}$ for the up-spin state, with a reduction of the total peak intensity in between. 
This observation supports  that the pair of parabolic surface bands in Fig. 2(a) originates from spin-split surface states.

\begin{figure}
\includegraphics[scale=1]{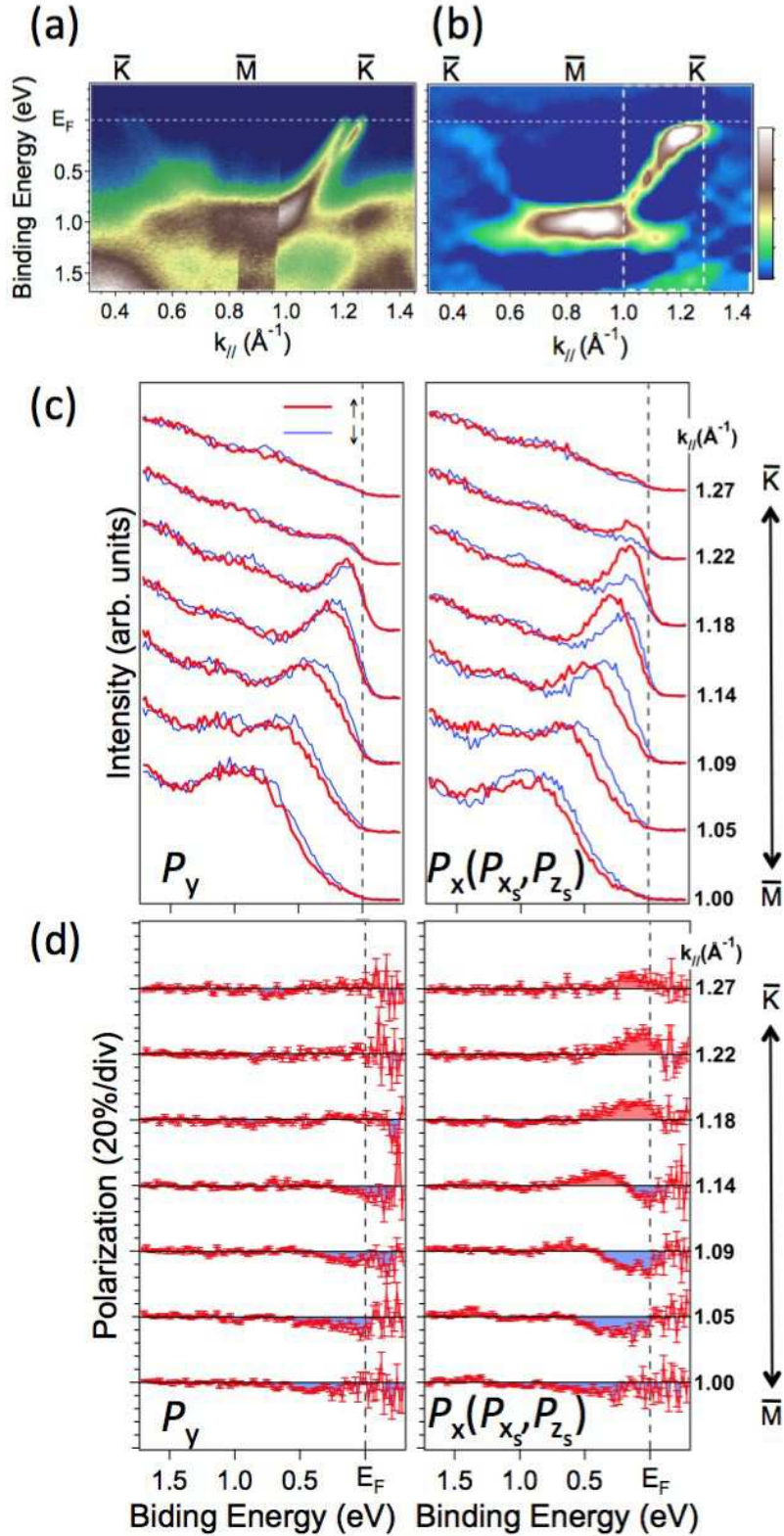}
\caption{\label{fig2:epsart} (Color online) $E-k_{\parallel}$ maps obtained by (a) spin-{\it integrated} and (b) spin-{\it resolved} ARPES measurements along the Au chains. The double surface state line-shape clearly observed in (a) is lost in (b) because of the lower energy and angular resolutions of SARPES. (c) SARPES spectra of the surface states around dashed area in (b) with $P_{\rm y}$ ($P_{\rm y_{\rm s}}$)  and $P_{\rm x}$ ($P_{\rm x_{\rm s}}$, $P_{\rm z_{\rm s}}$) polarization components in the spin-detector (sample) coordinate system and (d) its corresponding polarization spectra.}
\end{figure}

In order to confirm that the Rashba effect is the physical process behind this spin-splitting phenomenon, there must be a polarization reversal at  time-reversal symmetry points, such as the  $\bar{\rm M}$ point ( $k_{\parallel}$$\sim$0.81\AA$^{-1}$).
In Fig. 3 (a) transverse spin ($P_{\rm x}$) SARPES spectra of the surface states   are shown for an extended $k$-range (0.49$< $$k_{\parallel}$$<$1.35\AA$^{-1}$).
Above  $k_{\parallel}$=1.05\AA$^{-1}$, spin-split surface states dispersing towards $E_F$ are clearly observed. For  wave vectors below 1.00 \AA$^{-1}$ the spin splitting decreases and becomes practically zero as it approaches towards to the bottom of the parabola, the point  $\bar{\rm M}$  of the second SBZ. Crossing the $\bar{\rm M}$ point, the energy shift between up- and down-spin states appears again and the binding energy of up-spin state becomes now smaller than that of down-spin state. In other words, a polarization reversal at $\bar{\rm M}$ point is observed.
This polarization reversal shows clearly up in the angle resolved polarization map of Fig.3 (b).
These observations give robust evidence that the investigated one-dimensional, metallic double surface states become spin split by the Rashba effect. This is in agreement with the previous work obtained from first-principles calculations \cite{DSP04} and recent high-resolution ARPES measurement on Si(553)-Au\cite{Barke06}.

\begin{figure}
\includegraphics[scale=0.8]{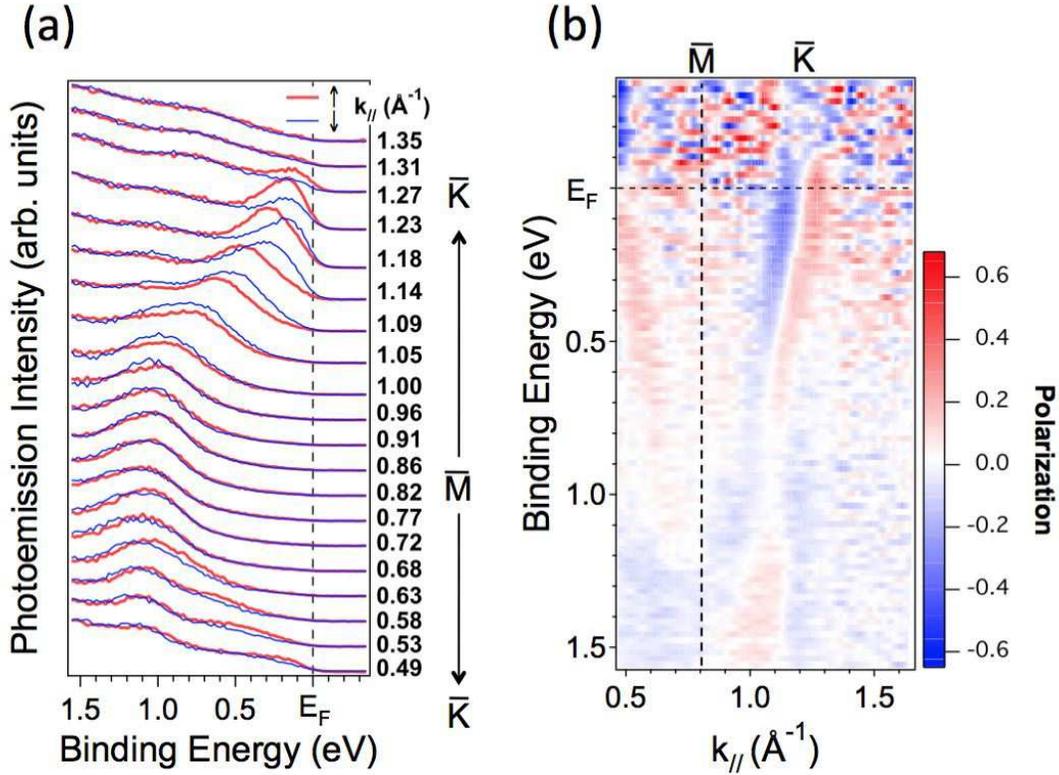}
\caption{\label{fig3:epsart}(Color) (a) SARPES spectra of the $P_{\rm x}$ ($P_{\rm x_{\rm s}}$, $P_{\rm z_{\rm s}}$) component in a wider $k$-range than  Fig.2(c). (b)  $E-k_{\parallel}$ polarization map of  $P_{\rm x}$ ($P_{\rm x_{\rm s}}$, $P_{\rm z_{\rm s}}$)  for  the same $k$-range as in (a).  }
\end{figure}

The resulting spin direction of the polarized surface states merits some further discussion. In an ideal two-dimensional Rashba case, as extracted from eq. 1, the spin direction must be orthogonal to both  the potential gradient, $\nabla V$  (surface normal direction) and to the electron momentum, $\overrightarrow{p}$, so that the spins are contained within the surface plane and  point tangential to the Fermi surface. Reducing the dimensionality to a one-dimensional case should result in a defined direction along x$_{\rm s}$  for $\overrightarrow{p}$, so that the spin direction would have to be perpendicular to the latter, i.e. have a  y$_{\rm s}$ (longitudinal) direction. However, what we observe in Fig.  2(c) and 2(d), is that the spin polarization is larger along the transverse direction ($P_{\rm x}$ (=$P_{\rm x_{\rm s}}$, $P_{\rm z_{\rm s}}$)) compared to the longitudinal one ($P_{\rm y}$(=$P_{\rm y_{\rm s}}$)). According to eq. 1,  is most unlikely that the spin should  point along the x$_{\rm s}$-direction, that is,  parallel to $\overrightarrow{p}$. Having in mind  that $P_{\rm x}$ is a mixture of $P_{\rm x_{\rm s}}$ and $P_{\rm z_{\rm s}}$, it  is  straightforward to conclude that  this large transverse spin component must be due to an out-of-plane spin component ($P_{\rm z_{\rm s}}$) projected onto the x-direction as illustrated in Fig. 1. The polar angle for the surface states observation in our SARPES measurements is around 25$^{\circ}$ ($\sim$$k_{\parallel}$=1.1\AA$^{-1}$), so that a  $P_{\rm z_{\rm s}}$ component projected onto the x-direction would yield about 50\%  of the total signal, which is therefore observable by our spin $P_{\rm x}$ channels.

To understand the origin of this $P_{\rm z_{\rm s}}$ component we have performed first-principles calculations on a Si(557)-Au repeated slab model\cite{DSP04} depicted in the inset of Fig. 4. 
Our numerical results stem from DFT calculations within the local density approximation with the all-electron full-potential linearized augmented plane wave (FLAPW) method. The spin splitting and spin structure have been investigated by switching on and off the spin-orbit coupling terms. 

\begin{figure}
\includegraphics[scale=0.6]{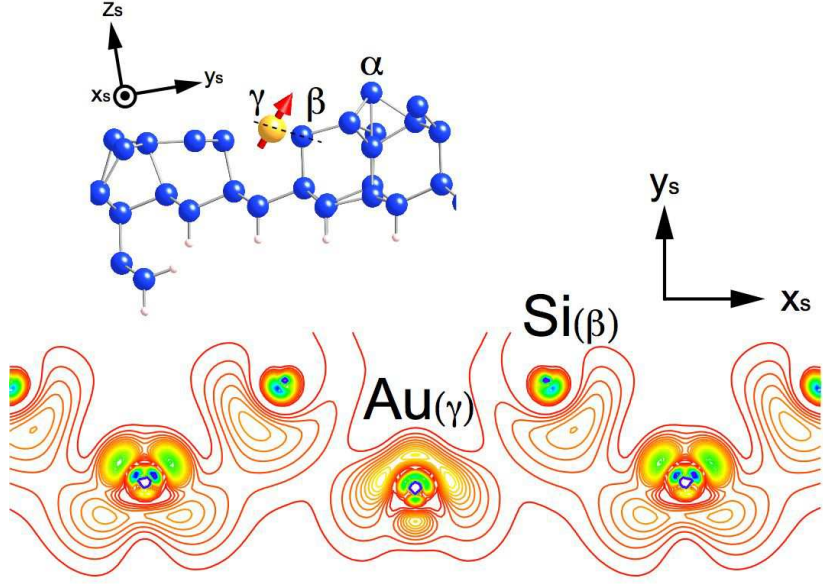}
\caption{\label{fig4:epsart} (Color online) Calculated charge distribution of the Au-Si chain structure in the plane determined by Au ($\gamma$) and Si ($\beta$) atoms (dashed line in the upper inset)  around $\bar{\rm K}$ and $E_{F}$. A side view of the model used in the calculations is shown in the  inset. A considerable large anisotropy for the charge distribution along the y$_{\rm s}$ direction (perpendicular to the chain (x$_{\rm s}$)) is found at the Au atoms which  is believed to cause the out-of-plane spin component (large red arrow in the model).}
\end{figure}

From  these calculations we could fairly reproduce the surface state splitting  reported in Ref.\cite{DSP04}.
The wave function character of these surface states corresponds to hybridized Au 5$d$ and 6$p$ orbitals, which is consistent with electronic states originating from the Au chains and not from the Si step edge structure. But this does not yet explain the existence of the observed out-of-plane component. Instead, we have calculated  the charge density of the parabolic bands (the wavefunction squared, $\mid\psi_{{\bm k},n}\mid^2$) around the Au-1D chain. The  charge distribution of the plane formed by the  Au ($\gamma$ in Fig. 1) and the Si ($\beta$) atoms (dashed line in the inset of Fig. 4) close to the $\bar{\rm K}$ point and $E_F$ is illustrated in Fig. 4. It is remarkable to observe that adjacent Au atoms are inequivalent,  in resemblance to the $\times$2 periodicity  along chain direction found for Si adatoms ($\alpha$), as reported from diffraction experiments and which upholds a band folding scenario close to the  $\bar{\rm K}$ point.
Furthermore, the charge distribution surrounding  the Au atoms is quite  anisotropic  along y$_{\rm s}$-direction, most likely because of the zigzag-like structure of the Au ($\gamma$) and Si ($\beta$) chains.
In this way, the potential gradient should have considerable y$_{\rm s}$-component so that previously unexpected  $P_{\rm z_{\rm s}}$ spin polarization can arise and even become dominant. Indeed, this is illustrated by the red arrow in the inset of Fig. 4: the calculated spin direction is almost perpendicular to the Au($\gamma$)-Si($\beta$) plane, which is consistent with our  SARPES observations.

In conclusion, we have investigated the spin structure of the nearly degenerated double surface states of Si(557)-Au by means of high energy- and angular-resolution SARPES.
Clear Rashba type spin-split surface states are experimentally observed, which are corroborated by our first-principles calculations including relativistic effects.
A dominating out-of-plane spin component has been observed and qualitatively explained by our calculation which we attribute to the one-dimensional structure of the Au chains. 
These findings allows us to shed some light over the controversial nature of these surface states,  finally settling this matter as originating from a Rashba effect.
The demonstration of  locked, dominant  out-of-plane  spin polarization  in one-dimensional metallic surface states due to  restricted electron momentum will very likely stimulate  revitalized research in metallic one-dimensional systems  within the spintronics and theory communities.

\begin{acknowledgments}

The authors acknowledge A. Nishide for the help in the experiment, J. Lobo-Checa for proof reading the manuscript, and 
D. S$\acute{\rm a}$nchez-Portal and S. Riikonen  for sharing their sample structural information  obtained by  first principles calculation.
This work was partly supported by KAKENHI (19340078), Grant-in-Aid for Scientific Research (B) of Japan Society for the Promotion of Science. 
The experiment was performed under the PF Proposal No. 2008G561.
The theoretical work was partly supported by Grant-in-Aid for Scientific Research (\#19GS0207) from the Ministry of Education, Culture, Sports, Science and Technology of Japan.

\end{acknowledgments}

%



\begin{thebibliography}{}
\bibitem{SBZ}The definition of the surface Brillouin zone (SBZ) is followed by that in Ref. \cite{DSP04}, which is artificially halved the SBZ in Ref. \cite{DSP02} in which the superlattice formation by adatom is neglected.
\end{thebibliography}
\end{document}